\documentclass{aip-cp}
\usepackage[numbers]{natbib}
\usepackage{rotating}
\usepackage{graphicx}
\usepackage{array}
\usepackage[english]{babel}
\usepackage[autostyle]{csquotes}
\begin{document}
\def\refname{REFERENCES}

% Title portion
\title{SPECTROSCOPY OF STRINGY BLACK HOLE}

\author[aff1]{\.{I}zzet Sakall\i{}\corref{cor1}}%\noteref{note1,note2}}
\eaddress[url]{http://drsakalli.com}
\author[aff1]{G\"{u}lnihal Tokg\"{o}z}%\noteref{note2}}
\eaddress{gulnihal.tokgoz@emu.edu.tr}

\affil[aff1]{Physics Department, Faculty of Arts and Sciences, Eastern Mediterranean
University, Famagusta, North Cyprus, via Mersin 10, Turkey.}
%\affil[aff2]{Additional affiliations should be indicated by superscript numbers 2, 3, etc. as shown above.}
%\affil[aff3]{You would list an author's second affiliation here.}
\corresp[cor1]{izzet.sakalli@emu.edu.tr}
%\authornote[note1]{This is an example of first authornote.}
%\authornote[note2]{This is an example of second authornote.}
\maketitle

\begin{abstract}
The Maggiore's method, which evaluates the transition frequency that appears
in the adiabatic invariant from the highly damped modes, is used to
investigate the entropy/area spectra of the Garfinkle--Horowitz--Strominger
black hole (GHS-BH). We compute the resonance modes of the GHS-BH by using
the confined scalar waves having high azimuthal quantum number. Although the
area and entropy are characterized by the GHS-BH parameters, their
quantizations are shown to be independent of those parameters. However, both
spectra are equally spaced.
\end{abstract}

% Title portion

%\noteref{note1,note2}}
\eaddress[url]{http://drsakalli.com}

%\noteref{note2}}
\eaddress{gulnihal.tokgoz@emu.edu.tr}

\affil[aff1]{Physics Department, Faculty of Arts and Sciences, Eastern Mediterranean
University, Famagusta, North Cyprus, via Mersin 10, Turkey.} 
%\affil[aff2]{Additional affiliations should be indicated by superscript numbers 2, 3, etc. as shown above.}
%\affil[aff3]{You would list an author's second affiliation here.}
\corresp[cor1]{izzet.sakalli@emu.edu.tr} 
%\authornote[note1]{This is an example of first authornote.}
%\authornote[note2]{This is an example of second authornote.}

% Head 1

\section{INTRODUCTION}

The effects of gravity are very strong near the black holes (BHs) and
quantum effects could not be ignored near the event horizon. Quantum gravity
theory (QGT) \cite{izr1} seeks to describe gravity according to the
principles of quantum mechanics. One of the difficulties of formulating the
QGT is that quantum gravitational effects only appear at length scales near
the Planck scale, around $10^{-35}$ meter, a scale far smaller, and
equivalently far larger in energy, than those currently accessible by high
energy particle accelerators. Therefore, we physicists lack experimental
data. The onset of the QGT dates back to the seventies. Hawking \cite%
{izr2,izr3} and Bekenstein \cite{izr4,izr5,izr6,izr7,izr8} showed that a BH
can be considered as a quantum mechanical system rather than a classical
object.

The quantization of BHs was first proposed in the seminal works of
Bekenstein \cite{izr5,izr6} in which the quantization procedure is based on
the surface area of a BH that acts as a classical adiabatic invariant.
According to the Ehrenfest principle \cite{izr9}, any classical adiabatic
invariant should have a quantum entity with a discrete spectrum. Thus,
Bekenstein \cite{izr5,izr6} suggested that the area of a quantum BH should
have a discrete and equally spaced spectrum:

\begin{equation}
\mathcal{A}_{n}=\epsilon n\hbar \ \ \ \ (n=0,1,2,3.......),  \label{ps01}
\end{equation}

where $\epsilon $ is an unknown \textquotedblleft fudge\textquotedblright\
factor \cite{izr10}\ and $\xi $\ is of the order of unity. One can
immediately deduce from Eq. (\ref{ps01}) that the minimum increase in the
area should be $\Delta \mathcal{A}_{\min }=\epsilon \hbar $. Moreover,
Bekenstein conjectured that for the family Schwarzschild BHs (including the
Kerr-Newman BH) the value of $\epsilon $ is $8\pi $ \cite{izr6}. In the
sequel, various methods were suggested to study the area spectrum of the BHs
and to determine the value of the coefficient $\epsilon $ (for the topical
review, a reader may refer to \cite{izr11} and references therein). Among
the methods proposed, Maggiore's method (MM) \cite{izr12} is the one that
its result perfectly fits with the Bekenstein's conjecture. In the MM, the
adiabatic invariant quantity ($I_{adb}$) is given by \cite%
{izr13,izr14,izr15,izr16}

\begin{equation}
I_{adb}=\hbar \int \frac{TdS}{\Delta \omega },  \label{ps02}
\end{equation}

in which $\Delta \omega =\omega _{n+1}-\omega _{n}$ stands for the
transition frequency between the subsequent levels of a BH with temperature (%
$T$) and entropy ($S$). On the other hand, according to the Bohr--Sommerfeld
quantization rule, $I_{adb}$\ is quantized as $I_{adb}\simeq n\hbar $ as $%
n\rightarrow \infty $. To obtain $\Delta \omega $, Maggiore \cite{izr12}
embraced the BH as a damped harmonic oscillator having a characteristic
frequency in the form of $\omega =\left( \omega _{R}^{2}+\omega
_{I}^{2}\right) ^{\frac{1}{2}}$: $\omega _{R}$\ and $\omega _{I}$\ are the
real and imaginary parts of the frequency, respectively. For the ultrahigh
dampings ($n\rightarrow \infty $), $\omega _{I}\gg \omega _{R}$ and
therefore $\Delta \omega \simeq \Delta \omega _{I}$. Hod \cite{izr17,izr18}
was the first physicist who considered the quasinormal modes (QNMs) or the
so-called ringing modes for computing $\Delta \omega $ and hence obtaining $%
I_{adb}$. Then, many studies have used the MM to test the Bekenstein's
conjecture for various BH solutions (see for instance \cite%
{izr19,izr20,izr21,izr22,izr23,izr24,izr26,izr27,izr28,izr29}). Today, there
are several methods to compute the QNMs, such as the WKB method, the phase
integral method, continued fractions and direct integrations of the wave
equation in the frequency domain \cite{izr292}.

In the present study, we study the entropy/area quantization (spectroscopy)
of the GHS-BH \cite{izr30}. GHS-BH is a solution to the low-energy limit of
the string theory whose its action is obtained when the Einstein--Maxwell
theory is expanded to involve a dilaton field $\phi $. That is why physical
properties of the GHS-BH is significantly different from the Reissner-Nordstr%
\"{o}m BH. The spectroscopy problem of the GHS-BH was first studied by Wei
et al. \cite{izr31}. In the adiabatic invariant quantity (\ref{ps02}), they
used the ordinary QNMs of Chen and Jing \cite{izr32} who obtained the
associated QNMs by using the monodromy method \cite{izr292}. Thus, it was
shown that GHS-BH has an equidistant area spectrum at the high frequency
modes. Later on, Sakalli and G\"{u}lnihal \cite{izr34} reconsidered the
problem of GHS-BH spectroscopy. However, instead of the ordinary QNMs they
computed the "\textit{boxed QNMs}" which are the characteristic resonance
spectra of the \textit{scalar clouds}. For that purpose, it was assumed that
there exists a \textit{mirror or confining cavity} surrounding the GHS-BH
which is placed at a constant radial coordinate with a radius $r_{m}$, which
is very close to the horizon. The scalar field is imposed to be vanished at
the mirror's location ($r_{m}$), which requires to use both the Dirichlet
and Neumann conditions. In that scenario, the radial wave equation was
studied around the near horizon region. Now, we want to reconsider the same
problem with another scenario. As seen in the following sections, the
effective potential generated from the massless Klein-Gordon equation (KGE)
performs a \textit{barrier peak (BP)} for the propagating scalar waves when
the azimuthal quantum number $l$ gets high values. As a result, the scalar
waves are confined between the horizon and the BP. This will yield
characteristic \textit{resonance modes (RMs)} of the confined scalar fields
in the GHS-BH geometry. To this end, the scalar field is imposed to be
terminated at the BP and to be purely ingoing wave at the horizon. In fact,
our method is similar to researches \cite{izr35,izr36,izr37,izr38} that are
mainly inspired from the studies of \cite{izr39,izr40} in which the QNMs are
computed using the poles of the scattering amplitude in the Born
approximation. After reducing the radial KGE to the one-dimensional Schr\"{o}%
dinger equation \cite{izr41}, we show that it becomes a confluent
hypergeometric differential equation \cite{izr42} near the horizon. Imposing
the relevant boundary condition and then using the pole feature of the Gamma
function, we obtain the RMs of the GHS-BH. Using the highly damping RMs in
the MM, we get the entropy/area spectra of the GHS-BH.

The paper is organized as follows. In Sec. 2, we introduce the GHS-BH metric
and study the massless KGE on this geometry. We also show how the radial
equation reduces to a one-dimensional Schr\"{o}dinger like wave equation.
Sec. 3 is devoted to the computation of the RMs of the GHS-BH. To this end,
we show how the Schr\"{o}dinger like wave equation reduces to a confluent
hypergeometric differential equation. Then, we apply the MM to obtain the
entropy/area spectra of the GHS-BH. We present our conclusions in Sec. 4.
(Throughout this paper, we use the unit of $c=G=k_{B}=1$).

\section{KGE ON GHS-BH GEOMETRY}

GHS-BH is the solution to the four-dimensional Einstein-Maxwell-dilaton
low-energy action \cite{izr30}. It has the following static and spherically
symmetric metric: 
\begin{equation}
ds^{2}=-f(r)dt^{2}+f(r)^{-1}dr^{2}+g(r)d\Omega ^{2},  \label{ps1}
\end{equation}

where the metric functions are given by

\begin{equation}
f(r)=\frac{r-2M}{r},  \label{ps2}
\end{equation}

\begin{equation}
g(r)=r^{2}-2ar,  \label{ps3}
\end{equation}

and

\begin{equation}
a=\frac{Q^{2}}{2M}e^{-2\phi _{0}}.  \label{ps4}
\end{equation}

Physical quantities $Q$, $M,$ and $\phi _{0}$ denote the magnetic charge,
mass, and the asymptotic value (constant) of the dilaton field,
respectively. $r_{h}=2M$ corresponds to the event horizon of the GHS-BH. It
is worth noting that in the GHS-BH geometry the dilaton field reads

\begin{equation}
e^{-2\phi }=\left( 1-\frac{2a}{r}\right) e^{-2\phi _{0}}.  \label{ps5}
\end{equation}

On the other hand, the Maxwell field is given by

\begin{equation}
F=Q\sin \theta d\theta \wedge d\varphi .  \label{ps6}
\end{equation}

One can also get the electric charge case via the following duality
transformations: 
\begin{equation}
\tilde{F}_{\mu \nu }=\frac{e^{-2\phi }}{2}\epsilon _{\mu \nu }^{\alpha \beta
}F_{\alpha \beta }, \ \ \ \ \ \ \ \ and \ \ \ \ \ \ \ \ \ \ \phi \rightarrow
-\phi .  \label{ps7}
\end{equation}

The surface gravities \cite{izr43} of the GHS-BH and the Schwarzschild BH
are the same:

\begin{equation}
\kappa =\lim_{r\rightarrow r_{h}}\sqrt{-\frac{1}{2}\nabla _{\mu }\chi _{\nu
}\nabla ^{\mu }\chi ^{\nu }}=\left. \frac{1}{2}f^{\prime }(r)\right\vert
_{r=r_{h}}=\frac{1}{4M},  \label{ps8}
\end{equation}

where $\chi ^{\nu }=[1,0,0,0]$ is the timelike Killing vector and the prime
symbol denotes the differentiation with respect to $r$. So, one can easily
obtain the Hawking temperature of the GHS-BH as

\begin{equation}
T_{H}=\frac{\hbar \kappa }{2\pi }=\frac{\hbar }{8\pi M}.  \label{ps9}
\end{equation}

On the other hand, the areal radii of the GHS-BH and the Schwarzschild BH
are different. Thus, the area and entropy of the GHS-BH are not same with
the Schwarzschild BH's ones:

\begin{equation}
S=\frac{\mathcal{A}}{4\hbar }=\frac{\pi (r_{+}-2a)r_{+}}{\hbar }.
\label{ps10}
\end{equation}

When $a=M$ (the extremal charge case: $Q=\sqrt{2}Me^{\phi _{0}}$), the
GHS-BH's area (and hence its entropy) vanishes. In fact, this extremal
GHS-BH is indeed a naked singularity. Moreover, for having physical entropy $%
S\geq 0$, it is necessary to have $a\leq M.$ The singularity of GHS-BH is
null (unlike to the timelike singularity of Reissner-Nordstr\"{o}m BH) and
therefore outward radial null geodesics do not hit it (a reader may refer to 
\cite{izr44}). The first law of thermodynamics for the GHS-BH takes the
following form: $T_{H}dS=dM-UdQ$. Here, $U$ is the electric potential
defined on the horizon: $U=aQ^{-1}$.

The massless KGE for the scalar field $\Psi $ is given by

\begin{equation}
\left( \sqrt{\left\vert g\right\vert }\right) ^{-1}\partial _{\nu }(\sqrt{%
\left\vert g\right\vert }g^{\mu \nu }\partial _{\mu }\Psi )=0.  \label{ps11}
\end{equation}

We shall use the following ansatz for the scalar field $\Psi $:

\begin{equation}
\Psi =g(r)^{-1/2}H(r)e^{i\frac{\omega }{\hbar }t}Y_{l}^{m}(\theta ,\varphi
), \ \ Re(\omega )>0,  \label{ps12}
\end{equation}

where $\omega \ $is the frequency of the propagating scalar wave. $%
Y_{l}^{m}(\theta ,\varphi )$ represents the spheroidal harmonics with the
eigenvalue $-l(l+1)$ and magnetic quantum number $m,$ respectively. Here, $l$
is the azimuthal quantum number, which is a type of quantum number defined
for an orbital which determines its orbital angular momentum and also
describes the shape of the orbital of a particle within the associated
geometry. It is worth noting that orbitals can take even more complex shapes
according to the higher values of $l$. A spherical orbit ($l=0$) can be
oriented in space in only one way. However orbital that has polar or
cloverleaf shapes can point in different directions. To describe the
orientation in space of a particular orbital, that is why one always needs
the magnetic quantum number $m.$

After some algebra, the radial equation becomes

\begin{equation}
\left( -\partial _{r^{\ast }}^{2}+V\right) H(r)-\left( \frac{\omega }{\hbar }%
\right) ^{2}H(r)=0,  \label{ps13}
\end{equation}

which is a one-dimensional Schr\"{o}dinger type differential equation \cite%
{izr41}. $r^{\ast }$ represents the tortoise coordinate, which is defined as

\begin{equation}
r^{\ast }=\int f(r)^{-1}dr.  \label{ps14}
\end{equation}

Evaluating integral of Eq. (\ref{ps14}), we get

\begin{equation}
r^{\ast }=r+r_{h}\ln \left( \frac{r-r_{h}}{r_{h}}\right) .  \label{ps15}
\end{equation}

On the other hand, one can express $r$ in terms of $r^{\ast }$ as follows

\begin{equation}
r=r_{h}\left[ 1+\Omega \left( z\right) \right] ,  \label{ps16}
\end{equation}

where $z=\exp \left( \frac{r^{\ast }-r_{h}}{r_{h}}\right) $ and $\Omega
\left( z\right) $ is the Lambert-W or the so-called omega function \cite%
{izr45}. The tortoise coordinate has the following limits:

\begin{equation}
\lim_{r\rightarrow r_{h}}r^{\ast }=-\infty , \ and\ \ \ \lim_{r\rightarrow
\infty }r^{\ast }=\infty .  \label{ps17}
\end{equation}

The effective potential $V(r)$\ seen in Eq. (\ref{ps13}) is found to be

\begin{equation}
V(r)=f(r)\left[ \frac{l(l+1)}{r\left( r-2a\right) }+\frac{r_{h}(r-a)}{%
r^{3}\left( r-2a\right) }-\frac{a^{2}f(r)}{\left[ r\left( r-2a\right) \right]
^{2}}\right] .  \label{ps18}
\end{equation}

Fig. (1) exhibits the plot of $V(r)$ versus $r^{\ast }$ for various values of%
$\ $the azimuthal quantum number $l$ with $M=1$ and $a=0.5$. It can be
deduced from Fig. (1) that when $l$--parameter takes higher values, the
effective potential tends to make a "\textit{BP}" at a specific point which
is among the event horizon and spatial infinity. This means that the scalar
waves having very high azimuthal quantum number ($l\gg 1$) that are not
sufficiently energetic could not pass that \textit{BP} and would be confined
in a small region. To obtain those RMs, as being discussed in the following
section, we shall make the analysis around the near horizon region of the
GHS-BH.

\begin{figure}[ht]
\includegraphics[scale=.4]{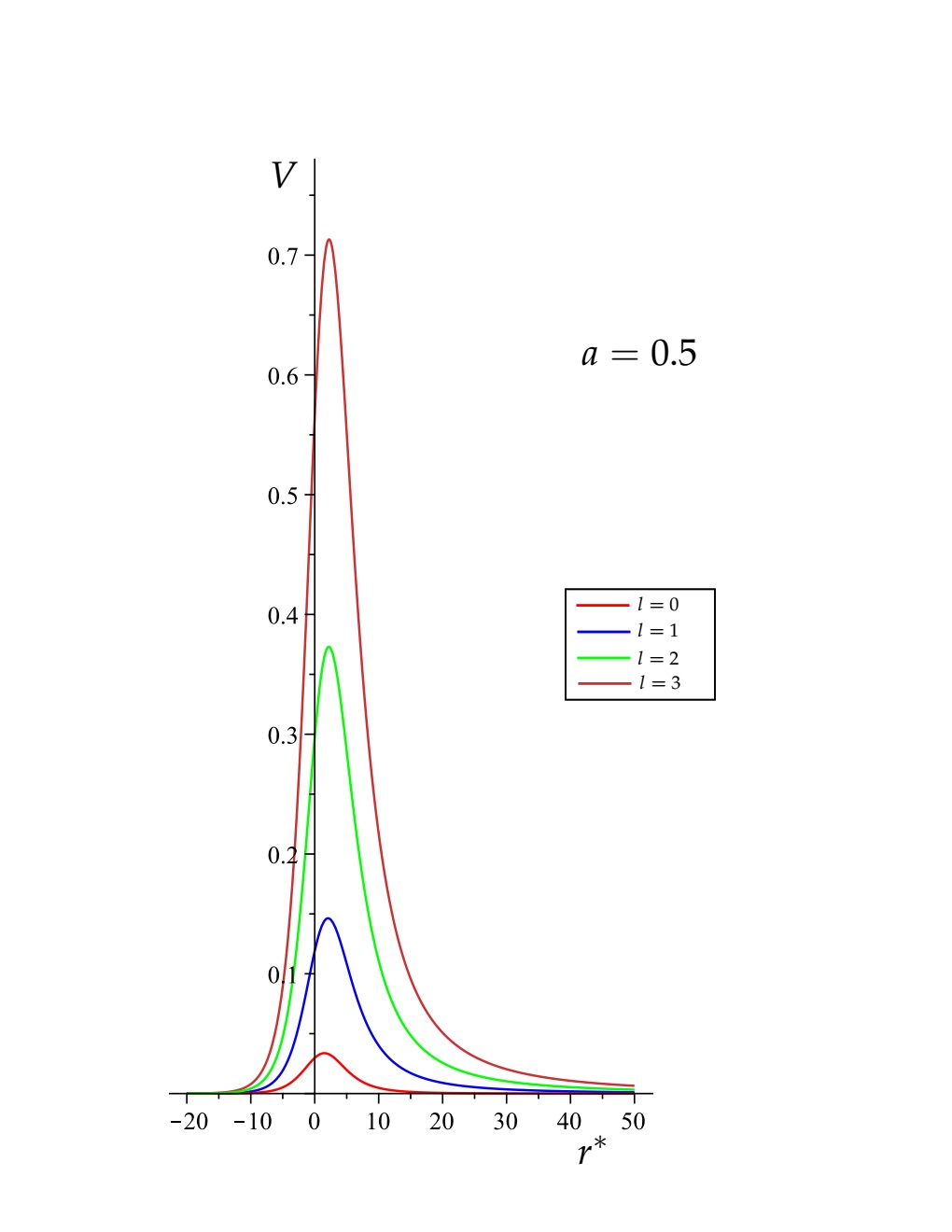} \vspace*{8pt}
\caption{Plot of $V$ versus $r^{\ast }$. The physical parameters are chosen
to be $M=1$ and $a=0.5 $. As $l$ gets bigger values, the potential barrier
between the horizon and spatial infinity exhaustively increases.}
\end{figure}

\section{SPECTROSCOPY ANALYSIS}

In principle, Eq. (\ref{ps13}) is solved for the QNMs with a particular set
of boundary conditions: purely ingoing wave at the horizon and purely
outgoing waves at the infinity. But unfortunately, Eq. (\ref{ps13}) cannot
be solved, analytically. Therefore to overcome this difficulty, one should
take the help of some approximate method. In this section, we shall follow a
particular method prescribed in \cite{izr35,izr36,izr37,izr38}, which
considers the wave dynamics in the vicinity of the event horizon. Since the
effective potential (\ref{ps17}) vanishes at the horizon ($r^{\ast
}\rightarrow -\infty $) and makes a barrier for the scalar waves with high
azimuthal quantum numbers ($l\gg 1$) at the intermediate region, therefore
the RMs are defined to be those for which one has purely ingoing plane wave
at the horizon and no wave at \textit{BP}'s location: the latter condition
is trivially satisfied. Namely, the relevant RMs should satisfy

\begin{equation}
\left. H(r)\right\vert _{RM}\sim \left\{ 
\begin{array}{c}
\:\:\:\:e^{\frac{i\omega r^{\ast }}{\hbar }}\:\:\: at\:\:\: r^{\ast
}\rightarrow -\infty \\ 
\\ 
0\:\:\: at\:\:\: \mathit{BP}%
\end{array}%
\right. .  \label{ps19}
\end{equation}

The metric function $f(r)$ can be expanded to series around the event
horizon as follows 
\begin{eqnarray}
f(r)\simeq f^{\prime }(r_{+})(r-r_{+})+\frac{f^{\prime \prime }(r_{+})}{2}%
(r-r_{+})^{2}+O[(r-r_{+})^{3}],  \nonumber \\
=2\kappa y++\frac{f^{\prime \prime }(r_{+})}{2}y^{2}+O\left( y^{3}\right) , 
\label{ps20}
\end{eqnarray}

where $y=r-r_{+}$. With this new variable, we apply the Taylor expansion
around $y=0$ to Eq. (\ref{ps18}) and obtain\ the near horizon form of the
effective potential as

\begin{equation}
V(y)\simeq 2\kappa y\left[ l(l+1)(C+Dy)+2\kappa (G+Hy)-2\kappa yN\right] ,
\label{ps21}
\end{equation}

where the parameters\ are\ given by

\begin{equation}
C=\frac{1}{z},\:\:\:D=-\frac{2x}{z^{2}},\:\:\:G=\frac{x}{z},\:\:\:H=-\frac{%
x^{2}+a^{2}}{z^{2}},\:\:\:N=\frac{a^{2}}{z^{2}},  \label{ps22}
\end{equation}

with

\begin{equation}
x =r_{+}-a, \:\:\:z=r_{+}(r_{+}-2a).  \label{ps23}
\end{equation}

Furthermore, the tortoise coordinate around the event horizon can be
expressed as

\begin{equation}
r^{\ast }\simeq \frac{1}{2\kappa }\ln y.  \label{ps24}
\end{equation}

Thus, near the horizon, the one-dimensional Schr\"{o}dinger equation (\ref%
{ps13}) behaves as follows 
\begin{equation}
-4\kappa ^{2}y^{2}\frac{d^{2}H(y)}{dy^{2}}-4\kappa ^{2}y\frac{dH(y)}{dy}+%
\left[ V(y)-\left( \frac{\omega }{\hbar }\right) ^{2}\right] H(y)=0.
\label{ps25}
\end{equation}

The above differential equation has two separate solutions, which can be
expressed in terms of the Whittaker functions \cite{izr42}. The solution can
be transformed to the CH functions \cite{izr42} and thus we have

\begin{equation}
H(y)\sim C_{1}y^{\frac{i\omega }{2\hbar \kappa }}{}M(\tilde{a},\tilde{b},%
\tilde{c}y)+C_{2}y^{\frac{i\omega }{2\hbar \kappa }}{}U(\tilde{a},\tilde{b},%
\tilde{c}y).  \label{ps26}
\end{equation}

The parameters of the above function are given by

\begin{eqnarray}
\tilde{a}& =-i\frac{\gamma }{\lambda \sqrt{\kappa }}+\frac{\tilde{b}}{2}, 
\nonumber \\
\tilde{b}& =1+i\frac{\omega }{\hbar \kappa },  \nonumber \\
\tilde{c}& =i\frac{\lambda }{2z\sqrt{\kappa }},  \label{ps27}
\end{eqnarray}

where

\begin{eqnarray}
\gamma &=&2\kappa x+l(l+1),  \nonumber \\
\lambda &=&4\sqrt{xl(l+1)+\kappa (z+3a^{2})}.  \label{ps28}
\end{eqnarray}

By using one of the transformations of the confluent hypergeometric
functions \cite{izr42}, we obtain the near horizon ($y\ll 1$) behavior of
the solution (\ref{ps26}) as 
\begin{equation}
H(y)\sim \left[ C_{1}+C_{2}\frac{\Gamma (1-\tilde{b})}{\Gamma (1+\tilde{a}-%
\tilde{b})}\right] y^{\frac{i\omega }{2\hbar \kappa }}+C_{2}\frac{\Gamma (%
\tilde{b}-1)}{\Gamma (\tilde{a})}y^{-\frac{i\omega }{2\hbar \kappa }},
\label{ps29}
\end{equation}

Since the RMs impose that the outgoing waves must spontaneously terminate at
the horizon, the second\ term must be vanished. This is possible with the
poles of the Gamma function of the denominator seen in the second term. In
short, if we set $\tilde{a}=-n$ $(n=0,1,2,...),$ the outgoing waves vanish
and hence we read the frequencies of the RMs of the GHS-BH. The result is
given by 

\begin{eqnarray}
\omega _{n} &=&\frac{\hbar \sqrt{\kappa }[2\kappa x+l(l+1)]}{2\sqrt{%
xl(l+1)+\kappa (x^{2}+2a^{2})}}+i(2n+1)\hbar \kappa ,  \nonumber \\
&\simeq &\frac{\hbar \sqrt{\kappa }}{2\sqrt{x}}l+i(2n+1)\hbar \kappa
,\:\:\:l\gg 1,  \label{ps30n}
\end{eqnarray}

where $n$ is called the overtone quantum (resonance) number \cite{izr46}.
For the highly excited states ($n\rightarrow \infty $ and therefore $\omega
_{I}\gg \omega _{R}$), we have 
\begin{equation}
\Delta \omega \approx \Delta \omega _{I}=2\kappa \hbar =4\pi T_{H}.
\label{ps31}
\end{equation}

Substituting this into Eq. (\ref{ps02}), we obtain

\begin{equation}
I_{adb}=\frac{S}{4\pi }\hbar .  \label{ps32}
\end{equation}

Recalling the Bohr-Sommerfeld quantization rule ($I_{adb}=\hbar n$), we find
the entropy spectrum as

\begin{equation}
S_{n}=4\pi n.  \label{ps33}
\end{equation}

Furthermore, since $S=\frac{\mathcal{A}}{4\hbar },$ we can also read the
area spectrum:

\begin{equation}
\mathcal{A}_{n}=16\pi \hbar n.  \label{ps34}
\end{equation}

Thus, the minimum area spacing becomes

\begin{equation}
\Delta \mathcal{A}_{\min }=16\pi \hbar .  \label{ps35}
\end{equation}

which represents that the entropy/area spectra of the GHS-BH are evenly
spaced. Same conclusion was obtained in the studies of \cite%
{izr35,izr36,izr37,izr38}. Moreover, our results support the Kothawala et
al.'s conjecture \cite{izr47} which claims that the BHs in Einstein's
theories should have equidistant area spectrum.

\section{CONCLUSION}

In this study, we have first studied the massless KGE on the GHSBH geometry.
Next, we have shown that the one-dimensional Schr\"{o}dinger type
differential equation (\ref{ps13}) can be obtained from the radial equation.
From figure (1), it is seen that for the high azimuthal quantum numbers ($%
l\gg 1$), the effective potential can form a \textit{BP} just beyond the
event horizon. In such a case, the scalar waves are confined between the
horizon and \textit{BP}. Then, we have applied the particular approximation
method \cite{izr35,izr36,izr37,izr38} for finding the characteristic
frequencies of the RMs at the near horizon region. We have shown that the
one-dimensional Schr\"{o}dinger type differential equation can be
approximated to a confluent hypergeometric differential equation. After some
straightforward computations, we have derived the RMs of the GHS-BH and in
sequel applied the MM for the highly damped RMs to find out the entropy/area
spectra of the GHS-BH. The obtained spectra are equally spaced and
independent of the physical parameters of the GHS-BH. As a final remark, our
calculations have revealed that the value of the dimensionless constant $%
\epsilon $ as $16\pi $. This result may be questioned since it is different
from the original result of Bekenstein: $\epsilon=8\pi $. However, as being
emphasized by Hod \cite{izr18}, rather than the value of $\epsilon$, the
uniform quantization of the area/entropy spectra has the utmost importance
in the subject of BH spectroscopy.

\section{ACKNOWLEDGMENTS}

We would like to congratulate and thank the Organizers and Editors for
making the 10th International Physics Conference of the Balkan Physical
Union -- BPU10 so worthwhile. % References

\end{document}